\documentclass[showpacs,aps,prb,amsmath,amssymb,superscriptaddress]{revtex4}
\usepackage{graphicx}
\usepackage{dcolumn}
\usepackage{bm}
\begin{document}

\baselineskip=0.55cm
\renewcommand{\thefigure}{\arabic{figure}}
\title{Dynamical properties of quasiparticles in a gapped graphene sheet}
\author{A. Qaiumzadeh}
\affiliation{Institute for Advanced Studies in Basic Sciences
(IASBS), Zanjan, 45195-1159, Iran}
\affiliation{School of Physics,
Institute for research in fundamental sciences, IPM
19395-5531 Tehran, Iran}
\author{F. K. Joibari}
\affiliation{School of Physics,
Institute for research in fundamental sciences, IPM
19395-5531 Tehran, Iran}
\author{Reza Asgari}
\affiliation{School of Physics,
Institute for research in fundamental sciences, IPM
19395-5531 Tehran, Iran}
\begin{abstract}
We present numerical calculations of the impact of charge
carriers-carriers interactions on the dynamical properties of quasiparticles
such as renormalized velocity and quasiparticle inelastic scattering lifetime
in a gapped graphene sheet. Our formalism is based on the many-body
$G_0W$-approximation for the self-energy. We present results for the many-body renormalized velocity suppression and the renormalization constant over a broad range of energy gap values. We find that the renormalized velocity is almost independence of the carrier densities at large density regime. We also show that the quasiparticle inelastic scattering lifetime decreases by increasing the gap value. Finally, we present results for the mean free path of charge carriers suppression over the energy gap values.

\end{abstract}
\pacs{71.10.Ay, 81.05.Uw, 71.45.Gm}
\maketitle

\section{Introduction}
The latest rival to succeed silicon's status is graphene, a single atomic layer of graphite make a truly tiny transistor to decrease the size and to improve the operational speed of the
electronic devices. Silicon lost it's brilliant electronic
properties in pieces smaller than about $10$nm and practically the
smallest silicon chips which has been used in silicon-based
electronics is $45$nm. Furthermore, silicon
has some limitations in speed of operations. These restrictions lead to
serious challenges for the Moore's law which states that the
number of transistors can be placed inexpensively on an integrated
circuit has increased exponentially, doubling approximately every
two years. This growth cannot be maintained forever and thus the
search is on to find and use new materials which may be able to produce
higher performance and better functionality.

The recent discovery~\cite{novoselov} of
graphene in $2004$, and its fabrication into a field-effect
transistor~\cite{novoselov}, has opened up a new field of physics and
offers exciting prospects for new electronic devices and
apparently possible to come over those aforementioned limitations.
Graphene has instructive and unique physics with special intriguing electronic
properties which has attracted remarkable attentions.~\cite{Barth}
First, the electronic properties of
graphene are improved in sizes less than $10$nm
. Second, the massless Dirac-like electrons move through graphene with
almost near-ballistic transport behavior with less resistance because back-scattering is suppressed. Third, graphene is itself a good thermal
conductor such that graphene's thermal conductivity is about
$\sim 5.3\times10^3$ W/mK at room temperature which is greater than the thermal
conductivity of carbon nanotubes.~\cite{thermal} Interestingly, the mobility of carriers in graphene is quite high
and it is about $10^5$ cm$^2$/Vs at room temperature.~\cite{morozov}
It is important to note that the highest electron mobility recorded on the semiconductor junction
{\rm H-Si}(111)-vacuum FET is $8\times10^3$cm$^2$/Vs at $4.2$ K
or the mobility of electrons in junction {\rm Si-SiO}${_2}$(100) MOSFET systems is $25\times10^3$cm$^2$/Vs at low
temperature~\cite{eng}, make graphene promising for different applications in devices.

Providing capability to control a type and density of charge
carriers by gate voltage or by the chemical doping~\cite{dop} made graphene
instructive for novel nano-electronic devices. However, a gapped semiconducting
behavior would be more suitable for electronic
applications.
There have been some proposed in literature for a gap generation in
graphene due to breaking of the sublattice symmetry by some
substrates (such as {\rm SiC}~\cite{3}, graphite~\cite{4} and boron
nitride\cite{5}), to adsorbe some molecules (such as water,
ammonia~\cite{6} and {\rm CrO}$_3$~\cite{cro}), spin-orbit
interaction~\cite{spin} and finite size effect.\cite{size} In case, we are interested to carry out the microscopic theory to calculate some physical quantities of gapped graphene.

Theoretical calculations of quasiparticle properties of electron in conventional two-dimensional electron liquid are performed within the framework of Landau's Fermi liquid theory~\cite{landau} whose key
ingredient is the quasiparticle concept and its interactions. As
applied to the electron liquid model this entails the calculation of
effective quasiparticle-quasiparticle interactions which enter the many-body
formalism allowing the calculation of various physical properties.
A number of calculations considered different variants of the
$G_0W$-approximation for the self-energy in two-dimensional electron gas\cite{em,yarlagadda_1994_2,em_bohm,em_dassarma,zhang,
asgari,asgari12} from which density, spin-polarization, and temperature
dependence of quasiparticle properties are obtained.

There is a mechanism for quasiparticle scattering against quasiparticles because they interact through the Coulomb interaction. This is an inelastic process and induced a finite lifetime of the quasiparticles. The carrier lifetime in an epitaxial graphene layers grown on {\rm SiC} wafers has been recently measured.~\cite{tau} Since experiments carried out their measurements on graphene placed on $\rm SiC$, we expect that graphene was gapped. The experimental measurements are relevant for understanding carrier intraband and interband scattering mechanisms in graphene and their impact on electronic and optical devises.~\cite{williams, gu}

In this paper we focus on the effect of energy gap on the
renormalized velocity, the
inelastic scattering lifetime of quasiparticles and the inelastic mean free path in gapped graphene sheets over the broad range of energy gap. Our formalism is based on the Landau-Fermi liquid theory incorporating the $G_0W$-approximation for the self-energy.
These quantities are related to some important physical properties of
both theoretical and practical applications such as the band structure of ARPES spectra~\cite{im}, the energy
dissipation rate of injected carriers~\cite{tau} and the width of the
quasiparticle spectral function.~\cite{martin}

The contents of the paper are described briefly as follows. In Section II we
discuss about our theoretical model which contains the effect of gap in the renormalized velocity of
quasiparticles and the inelastic scattering lifetime
$\tau_{in}$, of gapped graphene due to
electron-electron interactions by using $G_0W$-approximation. Our numerical results are given in Section III. Finally, Section V contains the summery and conclusions.

\section{Theoretical model}
Among the methods designed to deal with the intermediate correlation effects, of particular
interest for its physical appeal and elegance is Landau's phenomenological
theory~\cite{landau} dealing with low-lying excitations in a Fermi-liquid.
Landau called such single-particle excitations quasiparticles and postulated
a one-to-one correspondence between them and the excited states of a non-interacting
Fermi gas. He wrote the excitation energy of the Fermi-liquid in terms of the energies
of the quasiparticles and of their effective interaction. The quasiparticle-quasiparticle interaction function can
in turn be used to obtain various physical properties of the system and can be
parameterized in terms of experimentally measurable data. In this paper, we will compute the energy gap dependence of the renormalized velocity, renormalization constant and the inelastic scattering lifetime of quasiparticle in a gapped graphene sheet.

\subsection{Quasiparticle renormalized velocity}

The dynamics of quasiparticles in a gapless graphene are described
by two-dimensional (2D) massless Dirac Hamiltonian $\hat{H}=\hbar v {\bf \sigma}\cdot{\bf
k}$, with eigenvalues $\varepsilon_{s\bf k}=s\hbar v k$,
where $s=+(-)$ representing right- and left-handed helicity
or chirality for the electrons and holes, respectively. Note that chirality is the same as helicity for the massless particles. $v=10^{6}$ m$/$s is the Fermi velocity.
 As it has been shown before~\cite{velocity}, contrary to
conventional 2D electron systems, the interactions increase the
velocity of quasiparticles in graphene because of interband
exchange interactions and the difference between positive
and negative energy branches due to the chirality.

 The dynamics of quasiparticles in a gapped graphene are described by 2D massive Dirac Hamiltonian given by
$\hat{H}=\hbar v_F\sigma\cdot{\bf k}+mv^2\sigma_3$ with
eigenvalues $E_{s\bf k}=s\sqrt{(\hbar vk)^2+\Delta^2}$ where
$\Delta=mv^2$ is the gap energy. Due to massive term in the
Hamiltonian, the chirality differs from
the helicity and also the helicity is conserved but is frame dependence.

From the microscopic point of view, the quasiparticle energy can be calculated by solving the Dyson
equation,
\begin{eqnarray}
\delta\varepsilon_{s{\bf k}}^{QP}=\xi_{s{\bf k}}+\Re e[\delta
\Sigma_s^{ret}({\bf {\bf
k}},\omega)]|_{\omega=\delta\varepsilon_{s{\bf k}}^{QP}/\hbar},
\end{eqnarray}
where $\xi_{s{\bf k}}=E_{s{\bf k}}-E_{\rm F}$ is the energy of a quasiparticle relative to the Fermi energy. The Fermi wave vector in
graphene is given by $k_{\rm F}=(4\pi n/g_sg_v)^{1/2}$ where $g_s=g_v=2$
are spin and valley degeneracy, respectively. The Fermi energy of gapless graphene is $\varepsilon_{\rm F}=\hbar v k_{\rm F}$. The retarded self-energy of gapped graphene is $\Sigma_s^{ret}$ and we define $\delta
\Sigma_s^{ret}({\bf k},\omega)=\Sigma_s^{ret}({\bf
k},\omega)-\Sigma_s^{ret}(k_F,0)$. In the on-shell
approximation~\cite{Giuliani}, on the other hand, the above equation must be solved by setting
$\omega=\xi_{s{\bf k}}/\hbar$.

In the $G_0W$- approximation~\cite{Giuliani}, the
self-energy of gapped graphene at finite temperature ($\beta=1/(k_BT)$) is given by
\begin{eqnarray}
\Sigma_s({\bf{k}},i\omega_n)&=&-\frac{1}{\beta}\sum_{s'}\int\frac{d^2{\bf
q}}{(2\pi)^2}F^{ss'}({\bf
k,k+q})\\&\times&\sum_{m=-\infty}^{+\infty}W({\bf
q},i\Omega_m)G^{(0)}_{s'}({\bf k+q},i\omega_n+i\Omega_m)\nonumber,
\end{eqnarray}
where the dynamic screened effective interaction is $W({\bf q},i\Omega_m)=V_q/\epsilon(q,i\Omega_m)$ and $\epsilon(q,i\Omega_m)$ is the dynamical
dielectric function and the bare Coulomb
interaction is $V_q=2\pi e^2/\kappa q$ where $\kappa$ is the averaged background dielectric constant of graphene is placed on a substrate. $G_s^{(0)}(q,i\Omega_m)=1/(i\Omega_m-\xi_{s{\bf k}}/\hbar)$ is
the standard noninteracting Green's function.
The overlap
function for gapped graphene $F^{ss'}({\bf k,k+q})$, is given by~\cite{alireza}
\begin{eqnarray}
F^{ss'}({\bf k,k+q})= \frac{1}{2}(1+ss'\frac{\hbar^2 v^2{\bf
k}\cdot({\bf k}+{\bf q})+\Delta^2}{E_{\bf k}E_{\bf k+q}})~.
\end{eqnarray}
To evaluate of the
zero temperature retarded self-energy, we decompose the self-energy into the line which is purely a real function and residue
contributions, $\Sigma^{ret}_s({\bf
k},\omega)=\Sigma^{line}_s({\bf k}, \omega)+\Sigma^{res}_s({\bf
k},\omega)$, where $\Sigma^{line}$ is obtained by performing the
analytic continuation before summing over the Matsubara
frequencies, and $\Sigma^{res}$ is the correction which must be
taken into account in the total self-energy,
\begin{eqnarray}
\Sigma^{line}_s({\bf k}, \omega)&=&-\sum_{s'}\int\frac{d^2{\bf
q}}{(2\pi)^2}V_qF^{ss'}({\bf
k,k+q})\\&\times&\int_{-\infty}^{\infty}\frac{d\Omega}{2\pi}\frac{1}{\epsilon({\bf
q},i\Omega)}\frac{1}{\omega+i\Omega-\xi_{s'}({\bf
k+q})/\hbar},\nonumber
\end{eqnarray}
and
\begin{eqnarray}\label{res}
\Sigma^{res}_s({\bf k}, \omega)&=&\sum_{s'}\int\frac{d^2{\bf
q}}{(2\pi)^2}\frac{V_q}{\epsilon({\bf q},\omega-\xi_{s'}({\bf
k+q})/\hbar)}F^{ss'}({\bf
k,k+q})\nonumber\\&\times&[\Theta(\omega-\xi_{s'}({\bf
k+q})/\hbar)-\Theta(-\xi_{s'}({\bf k+q})/\hbar)],
\end{eqnarray}

where the dynamic dielectric function is given by $\epsilon({\bf q},\omega)=1-V_q\chi^{(0)}(q,\omega)$ in the random phase approximation (RPA) and
$\chi^{(0)}(q,\omega)$ is the noninteracting polarization function for gapped
graphene. The noninteracting polarization function has been recently calculated on both
along the imaginary and real frequency axis~\cite{alireza, pyatkovskiy}. The noninteracting polarization function expressions along the real frequency axis~\cite{pyatkovskiy} are given in appendix A.

Note that
there are two independent parameters in the self-energy. One of them is the Fermi energy $E_{\rm F}$, and the other is
the dimensionless coupling constant
$\alpha_{gr}=g_sg_ve^2/\kappa\hbar v$. The coupling constant in
graphene depends only on the substrate dielectric constant
while in the conventional 2D electron systems the coupling constant is density dependent.
For graphene placed on {\rm SiC} or graphite substrates, the coupling constant is about
$\alpha_{gr}\simeq 1$.

 The quasiparticle
energy depends on the magnitude of ${\bf k}$ for the isotropic systems. Expanding
$\delta\varepsilon_{+k}^{QP}$ to first order in $k-k_{\rm F}$, we obtain $\delta\varepsilon_{+k}^{QP}\simeq\hbar v^*(k-k_{\rm F})$ which
effectively defines the renormalized velocity as $\hbar
v^*=d\delta\varepsilon_{+k}^{QP}/dk|_{k=k_{\rm F}}$. The
renormalized velocity in the Dyson scheme is thus given by
\begin{eqnarray}\label{dyson}
\frac{v^*}{v}=\frac{(1+\Delta^2)^{-1/2}+v^{-1}\partial_k\Re
e[\delta \Sigma_+^{ret}({\bf
k},\omega)]|_{\omega=0,k=k_{\rm F}}}{1-\partial_{\omega}\Re e[\delta
\Sigma_+^{ret}({\bf k},\omega)]|_{\omega=0,k=k_{\rm F}}}~.
\end{eqnarray}
In the on-shell approximation, on the other hand, the renormalized
velocity is given by
$v^*/v=(1+\Delta^2)^{-1/2}+v^{-1}\partial_k\Re e[\delta
\Sigma_+^{ret}({\bf
k},\omega)]|_{\omega=0,k=k_{\rm F}}+(1+\Delta^2)^{-1/2}\partial_{\omega}\Re
e[\delta \Sigma_+^{ret}({\bf k},\omega)]|_{\omega=0,k=k_{\rm F}}$. The
renormalized velocity in this approximation demonstrates qualitatively the same
behavior obtained by the Dyson equation, Eq.~(\ref{dyson}) but its
magnitude is larger than the one calculated within the Dyson scheme.~\cite{asgari}
 There is an ultraviolet divergence in the wave vector integrals of the line contribution in a continuum model formulated as discussed above.~\cite{velocity} We introduce an ultraviolet cutoff
for the wave vector integrals, $k_c=\Lambda k_F$ which is the order
of the inverse lattice spacing and $\Lambda$ is dimensionless
quantity. For definiteness we take $\Lambda=k_c/k_F$ to be such that $\pi (\Lambda k_F)^2=(2\pi)^2/{\cal A}_0$, where
${\cal A}_0=3\sqrt{3} a^2_0/2$ is the area of the unit cell in the
honeycomb lattice, with $a_0 \simeq 1.42$~\AA~the carbon-carbon
distance. With this choice, $\Lambda\simeq{(g
n^{-1}\sqrt{3}/9.09)^{1/2}} \times 10^2$, where $n$ is the
electron density in units of $10^{12}~{\rm cm}^{-2}$.

An important quantity in the Fermi-liquid theory is the
renormalization constant $Z$, defined as the square of the overlap
between the state of the system after adding (or removing) of an
electron with the Fermi wave vector and the ground-state of the system. The non-zero renormalization constant value is always smaller than the one
for the normal Fermi-liquid systems and can be calculated explicitly as
follow\cite{Giuliani}
\begin{eqnarray}
Z=\frac{1}{1-\partial_{\omega}\Re e[\delta
\Sigma_+^{ret}({\bf k},\omega)] |_{\omega=0,k=k_{\rm F}}}.
\end{eqnarray}

We will show that $Z$ is a finite number for gapped graphene and it confirms as well that the system is a Fermi-Liquid.

\subsection{Inelastic scattering lifetime}

In this subsection, we compute the inelastic scattering lifetime of quasiparticles due to
carriers-carriers interactions at zero temperature and disorder-free for gapped graphene sheets. This is obtained through the imaginary part of the self-energy~\cite{inelastic} when
the frequency evaluated at the
on-shell energy.
\begin{eqnarray}
\tau_{in}^{-1}({\bf k})=\Gamma_{in}({\bf k},{\xi_{+\bf
k}}/\hbar)=-\frac{2}{\hbar}\Im m \Sigma_+^{ret}({\bf k},\xi_{+\bf
k}/\hbar),
\end{eqnarray}
where $\Gamma_{in}({\bf k},{\xi_{s\bf k}}/\hbar)$ is the quantum
level broadening of the momentum with eigenstate $|s{\bf k}>$. It is worthwhile to note that
the expression of $\tau_{in}^{-1}$ is identical with a result obtained by the Fermi's golden rule summing the scattering rate of electron and hole contributions at
wave vector ${\bf k}$.~\cite{Giuliani} Note again that the total contribution of the imaginary part of the retarded
self-energy comes from the residue term both intra- and interband contributions, $\Im
m\Sigma^{ret}_+({\bf k}, \omega)=\Im m\Sigma^{res}_{intra}({\bf
k}, \omega)+\Im m\Sigma^{res}_{inter}({\bf k}, \omega)$. However,
the total contribution of the imaginary part of the retarded
self-energy evaluated at the on-shell energy comes only from intraband term, $\Im m
\Sigma_+^{ret}({\bf
k},\xi_{\bf k}/\hbar)=\Im m\Sigma^{res}_{intra}({\bf k}, \xi_{\bf k}/\hbar)$. We will discuss about that with more details in the appendix B and C.

We turn our attention to investigate the imaginary part of the retarded
self-energy with more details. By starting from Eq.~(\ref{res}), we end up to an expression for the imaginary part of self-energy which is given by,
\begin{eqnarray}\label{imself}
\Im m\Sigma^{ret}_+({\bf k}, \omega)&=&\Im
m\Sigma^{res}_{intra}({\bf k}, \omega)+\Im
m\Sigma^{res}_{inter}({\bf k},
\omega)\nonumber\\&=&\int\frac{d^2{\bf q}}{(2\pi)^2}V_q\Im
m[\epsilon^{-1}({\bf q},\omega-\xi_{+}({\bf
k+q})/\hbar)]F^{++}({\bf
k,k+q})\nonumber\\&\times&[\Theta(\omega-\xi_{+}({\bf
k+q})/\hbar)-\Theta(-\xi_{+}({\bf k+q})/\hbar)]\nonumber\\&+&
\int\frac{d^2{\bf q}}{(2\pi)^2}V_q\Im m[\epsilon^{-1}({\bf
q},\omega-\xi_{-}({\bf k+q})/\hbar)]F^{+-}({\bf
k,k+q})\nonumber\\&\times&[\Theta(\omega-\xi_{-}({\bf
k+q})/\hbar)-\Theta(-\xi_{-}({\bf k+q})/\hbar)].
\end{eqnarray}
where the imaginary part of the inverse dielectric function in RPA level is obtained by
\begin{equation}\label{dielectric}
\Im m[\epsilon^{-1}({\bf q},\omega)]=\frac{V_q\Im
m\chi^{(0)}({\bf q},\omega)}{[1-V_q\Re e\chi^{(0)}({\bf
q},\omega)]^2+[V_q\Im m\chi^{(0)}({\bf q},\omega)]^2}.
\end{equation}

It is worth to note that the plasmon contributions in the imaginary part of self-energy comes from the zero-solutions of denominator in Eq.~(\ref{dielectric}).

\section{Numerical results}

We turn to a presentation of our main numerical results. We present
some illustrative results for the quasiparticle dynamic properties such renormalized velocity, renormalization constant and inelastic scattering lifetime. All numerical data are calculated in the Dyson scheme at $\alpha=1$.

The Fermi liquid phenomenology of Dirac electrons in
gapless graphene~\cite{im, velocity} and conventional
2D electron liquid~\cite{asgari} have the same structure, since both systems are
isotropic and have a single circular Fermi surface. The strength of
interaction effects in a conventional 2D electron liquid increases with decreasing
carrier density. At low densities, the quasiparticle renormalization constant $Z$ is
small, the renormalized velocity is suppressed~\cite{asgari}, the charge
compressibility changes sign from positive to negative,
and the spin-susceptibility is strongly enhanced~\cite{asgari2}.
These effects emerge from an interplay between exchange interactions
and quantum fluctuations of charge and spin in the 2D electron liquid.

In the 2D massless electron graphene, on the other hand, it has been
shown~\cite{yafis,velocity,im} that interaction effects also
become noticeable with decreasing density, although more slowly,
the quasiparticle renormalization constant, $Z$ tends to larger values, that the
renormalized velocity is enhanced rather than suppressed, and that the
influence of interactions on the compressibility and the
spin-susceptibility changes sign. These qualitative differences
are due to exchange interactions between electrons near the Fermi
surface and electrons in the negative energy sea and to interband
contributions to Dirac electrons from charge and spin
fluctuations.

In this paper we have shown the results for gapped graphene which are determined values between the gapless graphene evaluated at $\triangle=0$ and the conventional 2D electron liquid where $\triangle\rightarrow \infty$.

In Fig.~1, we have plotted the renormalized
velocity as a function of carrier density for
the various energy gap. As a result, we see that the impact
of energy gap on quasiparticles velocity which is similar to the effect of
impurity to that on graphene\cite{impurity}. The renormalized velocity is almost density independent in
gapped graphene at large carrier densities. The renormalized velocity reduces dramatically by increasing the
energy gap especially in the low carrier densities. Importantly, the renormalized velocity becomes less than the bare velocity at large energy gap and low density values. It is physically accepted since the system tends to conventional 2D electron liquid by increasing the energy gap values. Note that in the conventional 2D electron systems, the renormalized velocity is suppressed by increasing the coupling constant or reducing the density.

We have shown the renormalization constant $Z$, as a function of
the energy gap in Fig.~2. The renormalization constant
enormously reduces by increasing the energy gap in mild densities, however
it decreases quite slowly in high densities.

 Fig.~3(a) is shown the absolute value of $\Im m\Sigma^{\rm \scriptstyle ret}_+({\bf k},\omega)$ as from Eq.~(\ref{imself}), evaluated at $\omega=\xi_{\bf k}/\hbar$. By increasing the gap value, this function takes a finite jump at the wave number of the plasmon dip and at large $\triangle$ values, a discontinuity appears. The discontinuity is peculiar to $2D$ electron liquid.~\cite{giuliani_quinn} It is absent in gapless graphene and starts to arise from the fact that the oscillator strength of the plasmon pole is non-zero at special $k$ value for gapped graphene.

Fig.~3(b) is clearly shown the behavior of the energy gap dependence of the inverse inelastic scattering lifetime. As it is argued in the Appendix B, the imaginary part of self-energy evaluated at the on-shell energy start from $\triangle-\sqrt{\varepsilon_{\rm F}^2+\triangle^2}$ and in case the results are truncated below that. The quasiparticle lifetime decreases by increasing the gap value and it is a clear difference between 2D massless Dirac electron and gapped graphene. Consequently, the inelastic scattering lifetime in graphene is always larger than the conventional 2D electron liquid. In the case of gapless graphene, scattering rate is
a smooth function because of the absence of both plasmon emission
and interband processes,\cite{inelastic} nevertheless with generating a
gap and increasing the amount of it, plasmon emission causes to arise a discontinuity in the
scattering time, similar to conventional 2D electron liquid.~\cite{asgari} We have thus two mechanisms for scattering of the quasiparticles. The excitation of electron-hole pairs which is dominant process at low wave vectors and the excitation of plasmon appears in a specific wave vector. We also see in Fig.~3(b) that the scattering rate is quite sensitive to the gap energy and the scattering rate increases by increasing the energy gap.

In Fig.~4, we have depicted the inelastic mean free path
$l_{in}({\bf k})=v^*\tau_{in}({\bf k})$, as a function of the on-shell energy for various
gap energies. To this purpose we multiplied the results of $\tau_{in}({\bf k})$ to a proper renormalized velocity. As a result the mean free path of a gapped graphene is
shorter than that obtained for gapless graphene. Furthermore, the massless graphene has larger $l_{in}$ and it decreases by increasing the energy gap values. Note that the typical value of energy gap due to breaking sublattice symmetry is $\triangle=10-100$ meV corresponding the inelastic mean free path is $l_{in}=20-50$ nm which implies that the system remains in the semi-ballistic
regime.~\cite{3,4,5,6}.

\section{Summery and concluding remarks}

In summary, we have studied the problem of the microscopic
calculation of the quasiparticle self-energy and many-body renormalized velocity suppression over the energy gap
in a gapped graphene. We have carried out calculations of both the real and the imaginary part of the
quasiparticle self-energy within $G_0W$-approximation. We have also presented results for the renormalized velocity
suppression and for the renormalization constant over a wide range of energy gap. We have shown that the renormalized velocity for a gapped graphene is almost independent of the carrier density at high density.
We have finally presented results for the quasiparticle inelastic scattering lifetime suppression over the energy gap and show that the mean free path of the charge carriers of a gapless graphene is larger than a gapped graphene one. In case the mean free path of charge carriers decrease by increasing the energy gap.

A possible role of correlations including the charge-density fluctuations beyond the Random Phase Approximation,
remains to be examined.

\begin{acknowledgments}

 R. A. would like to thank the International Center for Theoretical Physics, Trieste for its hospitality during the period when part of this work was carried out. A. Q is supported by IPM grant.
\end{acknowledgments}

\appendix
\section{The dynamic polarization function for a gapped graphene}
In this appendix we present the real and imaginary part of the
noninteracting polarization function for a gapped graphene, which is
calculated recently by Pyatkovskiy.~\cite{pyatkovskiy} The dynamic polarization
function for gapped graphene in the imaginary frequency axis is also
calculated by us in Ref. [\onlinecite{alireza}]. Importantly, the noninteracting polarization function along the imaginary frequency axis can be obtained by performing analytical continuation from real axis and those results are the same.~\cite{private}

First, by introducing some following notations,
\begin{eqnarray}
f(k,\omega)&=&\frac{g_sg_vk^2}{16\pi\sqrt{|\hbar^2v^2k^2-\hbar^2\omega^2|}},\nonumber\\
g_{\pm}&=&\frac{2E_{\rm F}\pm\hbar \omega}{\hbar v k},\nonumber\\
x_0&=&\sqrt{1+\frac{4\Delta^2}{\hbar^2v^2k^2-\hbar^2\omega^2}},\nonumber\\
G_<(x)&=&x\sqrt{x_0^2-x^2}-(2-x_0^2)\cos^{-1}(x/x_0),\nonumber\\
G_>(x)&=&x\sqrt{x^2-x_0^2}-(2-x_0^2)\cosh^{-1}(x/x_0),\nonumber\\
G_0(x)&=&x\sqrt{x^2-x_0^2}-(2-x_0^2)\sinh^{-1}(x/\sqrt{-x_0^2}),
\end{eqnarray}
the real part of noninteracting polarization function is given by,
\begin{eqnarray}
\Re e \chi^{(0)}(k,\omega)&=&-\frac{g_sg_vE_{\rm F}}{2\pi
v_F^2}+f(k,\omega)\times\left\{
\begin{array}{ll}
  0, & \hbox{1A} \\
  G_<(g_{-}), & \hbox{2A} \\
  G_<(g_{+})+G_<g_{-}), & \hbox{3A} \\
  G_<(g_{-})-G_<(g_{+}), & \hbox{4A} \\
  G_>(g_{+})-G_>(g_{-}), & \hbox{1B} \\
  G_>(g_{+}), & \hbox{2B} \\
  G_>(g_{+})-G_>(-g_{-}), & \hbox{3B} \\
  G_>(-g_{-})+G_>(g_{+}), & \hbox{4B} \\
  G_0(g_{+})-G_0(g_{-}), & \hbox{5B} \\
\end{array}
\right.
\end{eqnarray}
and the imaginary part of noninteracting polarization function is given
by,
\begin{eqnarray}
\Im m \chi^{(0)}(k,\omega)&=&f(k,\omega)\times \left\{
\begin{array}{ll}
    G_>(g_{+})-G_>(g_{-}), & \hbox{1A} \\
    G_>(g_{+}), & \hbox{2A} \\
    0, & \hbox{3A} \\
    0, & \hbox{4A} \\
    0, & \hbox{1B} \\
    -G_<(g_{-}), & \hbox{2B} \\
    \pi(2-x_0^2), & \hbox{3B} \\
    \pi(2-x_0^2), & \hbox{4B} \\
    0, & \hbox{5B} \\
\end{array}
\right.
\end{eqnarray}
with the followings regions in the $(k,\omega)$ space,
\begin{eqnarray}
\begin{array}{cc}
  1A & \hbar\omega<E_{\rm F}-\sqrt{\hbar^2v^2(k-k_{\rm F})^2+\Delta^2}, \\
  2A & |E_{\rm F}-\sqrt{\hbar^2v_F^2(k-k_{\rm F})^2+\Delta^2}|<\hbar\omega<-E_{\rm F}+\sqrt{\hbar^2v^2(k+k_{\rm F})^2+\Delta^2}, \\
  3A & \hbar\omega<-E_{\rm F}+\sqrt{\hbar^2v^2(k-k_{\rm F})^2+\Delta^2}, \\
  4A & -E_{\rm F}+\sqrt{\hbar^2v^2(k+k_{\rm F})^2+\Delta^2}<\hbar\omega<\hbar v k, \\
  1B & k<2k_{\rm F},~~~\sqrt{\hbar^2 v^2k^2+4\Delta^2}<\hbar\omega<E_{\rm F}+\sqrt{\hbar^2v^2(k-k_{\rm F})^2+\Delta^2},\\
  2B & E_{\rm F}+\sqrt{\hbar^2v^2(k-k_{\rm F})^2+\Delta^2}<\hbar\omega<E_{\rm F}+\sqrt{\hbar^2v^2(k+k_{\rm F})^2+\Delta^2}, \\
  3B & \hbar\omega>E_{\rm F}+\sqrt{\hbar^2v^2(k+k_{\rm F})^2+\Delta^2}, \\
  4B & k>2k_{\rm F},~~~\sqrt{\hbar^2 v^2k^2+4\Delta^2}<\hbar\omega<E_{\rm F}+\sqrt{\hbar^2v^2(k-k_{\rm F})^2+\Delta^2},  \\
  5B & \hbar v k<\hbar\omega<\sqrt{\hbar^2 v^2k^2+4\Delta^2}, \\
\end{array}
\end{eqnarray}

\section{The intraband contribution of self-energy}
Since we are interested in quasiparticle properties, we therefore need only $s=+$ contribution. Let us focus on the intraband contribution of the retarded self-energy. The second argument of the
dielectric function in Eq. (\ref{res}) ( by setting $\hbar=v=1$) is
\begin{eqnarray}
\omega-\xi_{+}({\bf
k+q})=\omega+E_{\rm F}-\sqrt{k^2+q^2+2kq\cos\phi+\Delta^2}~.
\end{eqnarray}
In this case, we change the variable $\phi$ and integrate it over
$y=\sqrt{k^2+q^2+2kq\cos\phi+\Delta^2}$.
Using the new variable, the intraband contribution of self-energy changes to
\begin{eqnarray}\label{intra}
\Sigma_{intra}^{res}({\bf
k},\omega)&=&\frac{e^2}{2\pi\kappa\sqrt{k^2+\Delta^2}}\int_0^{+\infty}dq\int_{\sqrt{(k-q)^2+\Delta^2}}
^{\sqrt{(k+q)^2+\Delta^2}}
\frac{dy}{\sqrt{4k^2q^2-(y^2-k^2-q^2-\Delta^2)^2}}\frac{(\sqrt{k^2+\Delta^2}+y)^2-q^2}
{\epsilon({\bf
q},\omega+E_{\rm F}-y)}\nonumber\\&\times&[\Theta(\omega+E_{\rm F}-y)-\Theta(E_{\rm F}-y)].
\end{eqnarray}
We can now simplify the $\Theta$-functions further in Eq.~(\ref{intra}) by considering the positive and negative regions of $\omega$ as follow
\begin{eqnarray}
1) &\omega&+E_{\rm F}-y>0 ~and~ ~E_{\rm F}-y<0: ~It~ implies~ that~ \omega>0~,\nonumber\\
2) &\omega&+E_{\rm F}-y<0 ~and~ ~E_{\rm F}-y>0 : ~It~ implies~ that~ \omega<0~.\nonumber
\end{eqnarray}

To consider the first case where $\omega>0$, the difference
between the two $\Theta$-functions in Eq.~(\ref{intra}) is equal to $+1$ if
\begin{eqnarray}
E_{\rm F}<y<\omega+E_{\rm F}~~and~~\sqrt{(k-q)^2+\Delta^2}<y<\sqrt{(k+q)^2+\Delta^2}~.
\end{eqnarray}
Now we do need to find the overlap between these two intervals. We simply end up to inequivalent conditions which are
$q>k-\sqrt{\omega^2+k_{\rm F}^2+2\omega\sqrt{k_{\rm F}^2+\Delta^2}}$,
$q<k+\sqrt{\omega^2+k_{\rm F}^2+2\omega\sqrt{k_{\rm F}^2+\Delta^2}}$ and
$q>k_{\rm F}-k$. Collecting everything together and using the fact that $q\geq0$, we
finally find
\begin{eqnarray}
\Sigma_{intra}^{res}({\bf
k},\omega>0)&=&\frac{e^2}{2\pi\kappa\sqrt{k^2+\Delta^2}}
\int_{max(0,k_{\rm F}-k,k-\sqrt{\omega^2+k_{\rm F}^2+2\omega\sqrt{k_{\rm F}^2+\Delta^2}})}
^{k+\sqrt{\omega^2+k_{\rm F}^2+2\omega\sqrt{k_{\rm F}^2+\Delta^2}}}
dq\int^{min(\omega+\sqrt{k_{\rm F}^2+\Delta^2},\sqrt{(k+q)^2+\Delta^2})}
_{max(\sqrt{k_{\rm F}^2+\Delta^2},\sqrt{(k-q)^2+\Delta^2})}dy
\nonumber\\&\times&\frac{(\sqrt{k^2+\Delta^2}+y)^2-q^2}{\epsilon({\bf
q},\omega+E_{\rm F}-y)\sqrt{4k^2q^2-(y^2-k^2-q^2-\Delta^2)^2}}
\end{eqnarray}
 By considering of the second case where $\omega<0$, the
difference between the two $\Theta$-functions in Eq.~(\ref{res}) is equal
to $-1$ if
\begin{eqnarray}
E_{\rm F}+\omega<y<E_{\rm F}~~and~~\sqrt{(k-q)^2+\Delta^2}<y<\sqrt{(k+q)^2+\Delta^2}
\end{eqnarray}
As what we did before, we calculate overlap between intervals and thus we find
$q>k-k_{\rm F}$ and $q<k+k_{\rm F}$,
$q>\sqrt{\omega^2+k_{\rm F}^2+2\omega\sqrt{k_{\rm F}^2+\Delta^2}}-k$. Putting
everything together and using the fact that $q\geq0$ we finally find
\begin{eqnarray}
\Sigma_{intra}^{res}({\bf
k},\Delta-E_{\rm F}<\omega<0)&=&-\frac{e^2}{2\pi\kappa\sqrt{k^2+\Delta^2}}
\int_{max(0,k-k_{\rm F},\sqrt{\omega^2+k_{\rm F}^2+2\omega\sqrt{k_{\rm F}^2+\Delta^2}}-k)}^{k+k_{\rm F}}
dq\nonumber\\&\times&\int^{min(\sqrt{k_{\rm F}^2+\Delta^2},\sqrt{(k+q)^2+\Delta^2})}_{max(0,\omega+\sqrt{k_{\rm F}^2+\Delta^2}
,\sqrt{(k-q)^2+\Delta^2})}dy\nonumber\\&\times&\frac{(\sqrt{k^2+\Delta^2}+y)^2-q^2}
{\epsilon({\bf
q},\omega+E_{\rm F}-y)\sqrt{4k^2q^2-(y^2-k^2-q^2-\Delta^2)^2}}
\end{eqnarray}
\begin{eqnarray}
\Sigma_{intra}^{res}({\bf
k},\omega<-\Delta-E_{\rm F})&=&-\frac{e^2}{2\pi\kappa\sqrt{k^2+\Delta^2}}
\int_{max(0,k-k_{\rm F})}^{k+k_{\rm F}}
dq\int^{min(\sqrt{k_{\rm F}^2+\Delta^2},\sqrt{(k+q)^2+\Delta^2})}_{max(0,\omega+\sqrt{k_{\rm F}^2+\Delta^2},\sqrt{(k-q)^2+\Delta^2})}dy
\nonumber\\&\times&\frac{(\sqrt{k^2+\Delta^2}+y)^2-q^2}
{\epsilon({\bf
q},\omega+E_{\rm F}-y)\sqrt{4k^2q^2-(y^2-k^2-q^2-\Delta^2)^2}}
\end{eqnarray}

The real and imaginary part of intraband contributions can be computed.

\section{The interband contribution of self-energy}

Now, we focus on the interband contribution of the retarded self-energy. The second argument of the
dielectric function in Eq.~(\ref{res}) is
\begin{eqnarray}
\omega-\xi_{-}({\bf
k+q})=\omega+E_{\rm F}+\sqrt{k^2+q^2+2kq\cos\phi+\Delta^2}~.
\end{eqnarray}
We change variable
$y=\sqrt{k^2+q^2+2kq\cos\phi+\Delta^2}$, then we find
\begin{eqnarray}
\Sigma_{inter}^{res}({\bf
k},\omega)&=&\frac{e^2}{2\pi\kappa\sqrt{k^2+\Delta^2}}\int_0^{+\infty}dq\int_{\sqrt{(k-q)^2+\Delta^2}}
^{\sqrt{(k+q)^2+\Delta^2}}
\frac{dy}{\sqrt{4k^2q^2-(y^2-k^2-q^2-\Delta^2)^2}}\frac{q^2-(y-\sqrt{k^2+\Delta^2})^2}
{\epsilon({\bf
q},\omega+E_{\rm F}+y)}\nonumber\\&\times&[\Theta(\omega+E_{\rm F}+y)-1].
\end{eqnarray}
Note that $\Sigma_{inter}^{res}$ can be non-zero if
$\omega+E_{\rm F}+y<0~~and~~y>0$. It means that $\omega<-E_{\rm F}$.
In this case the difference between the two $\Theta$-functions in
Eq.~(\ref{res}) becomes -1 if
\begin{eqnarray}
0<y<-(\omega+E_{\rm F})~~and~~\sqrt{(k-q)^2+\Delta^2}<y<\sqrt{(k+q)^2+\Delta^2}~.
\end{eqnarray}
Now we do need to find the overlap between these two
intervals. We end up to inequivalent conditions that
$q>k-\sqrt{\omega^2+k_{\rm F}^2+2\omega\sqrt{k_{\rm F}^2+\Delta^2}}$ and
$q<k+\sqrt{\omega^2+k_{\rm F}^2+2\omega\sqrt{k_{\rm F}^2+\Delta^2}}$. Putting
everything together and using the fact that $q\geq0$ we finally get
\begin{eqnarray}
\Sigma_{inter}^{res}({\bf
k},\omega<-E_{\rm F})&=&-\frac{e^2}{2\pi\kappa\sqrt{k^2+\Delta^2}}
\int_{max(0,k-\sqrt{\omega^2+k_{\rm F}^2+2\omega\sqrt{k_{\rm F}^2+\Delta^2}})}^{k+\sqrt{\omega^2+k_{\rm F}^2+2\omega\sqrt{k_{\rm F}^2+\Delta^2}}}
dq\int_{\sqrt{(k-q)^2+\Delta^2}}^{min(\sqrt{(k+q)^2+\Delta^2},-(\omega+\sqrt{k_{\rm F}^2+\Delta^2}))}dy\nonumber\\
&\times&\frac{q^2-(y-\sqrt{k^2+\Delta^2})^2}{\epsilon({\bf
q},\omega+\varepsilon_{\rm F}+y)\sqrt{4k^2q^2-(y^2-k^2-q^2-\Delta^2)^2}} .
\end{eqnarray}

If we want to calculate $\Im m\Sigma^{res}_+({\bf
k},\xi_+(\bf k))$ needed for computing the quasiparticle lifetime,
we will only need the intraband contribution of the self-energy since the interband contribution is zero.

\newpage

\begin{figure}[t]
\includegraphics[width=8cm]{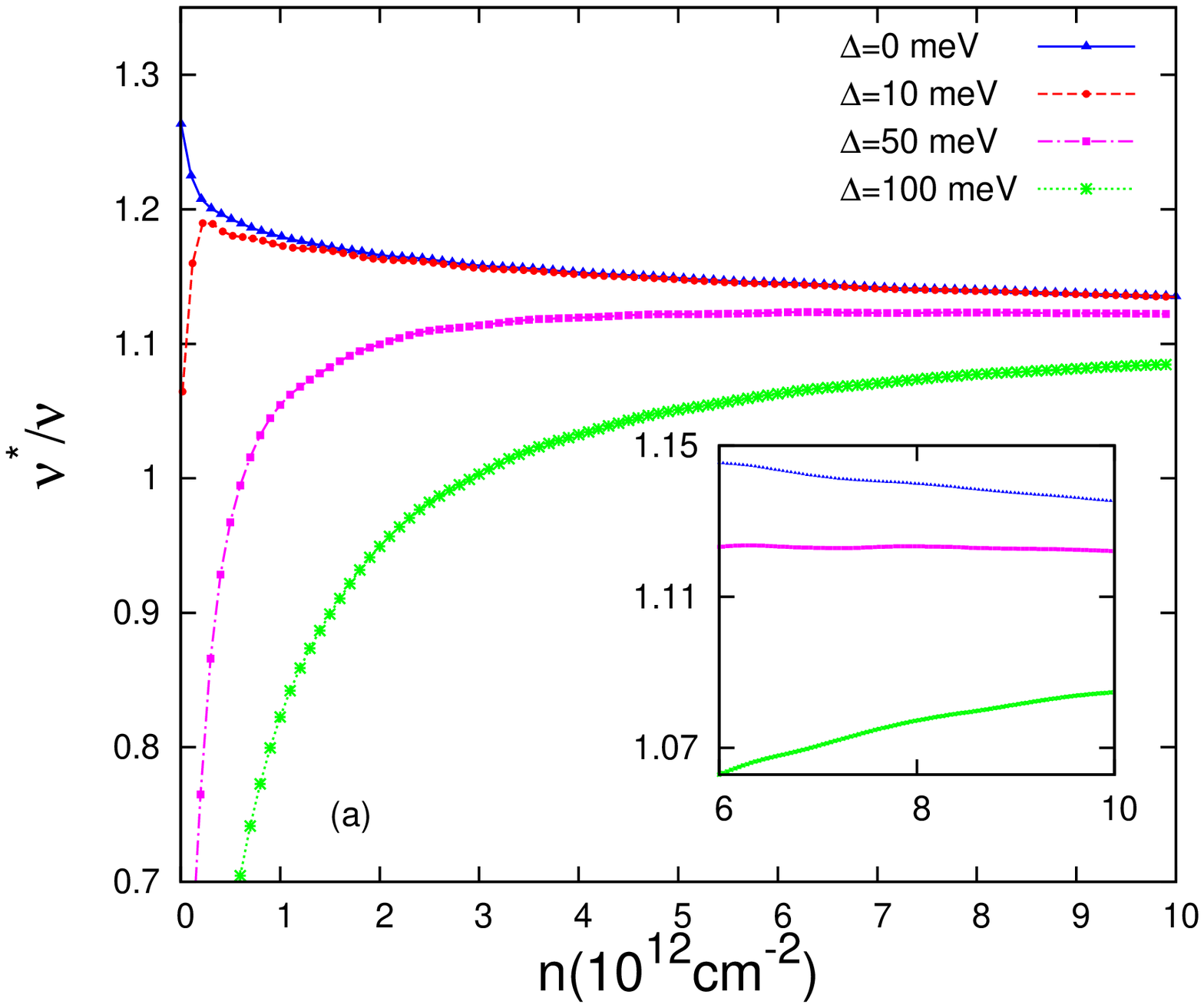}
\includegraphics[width=8cm]{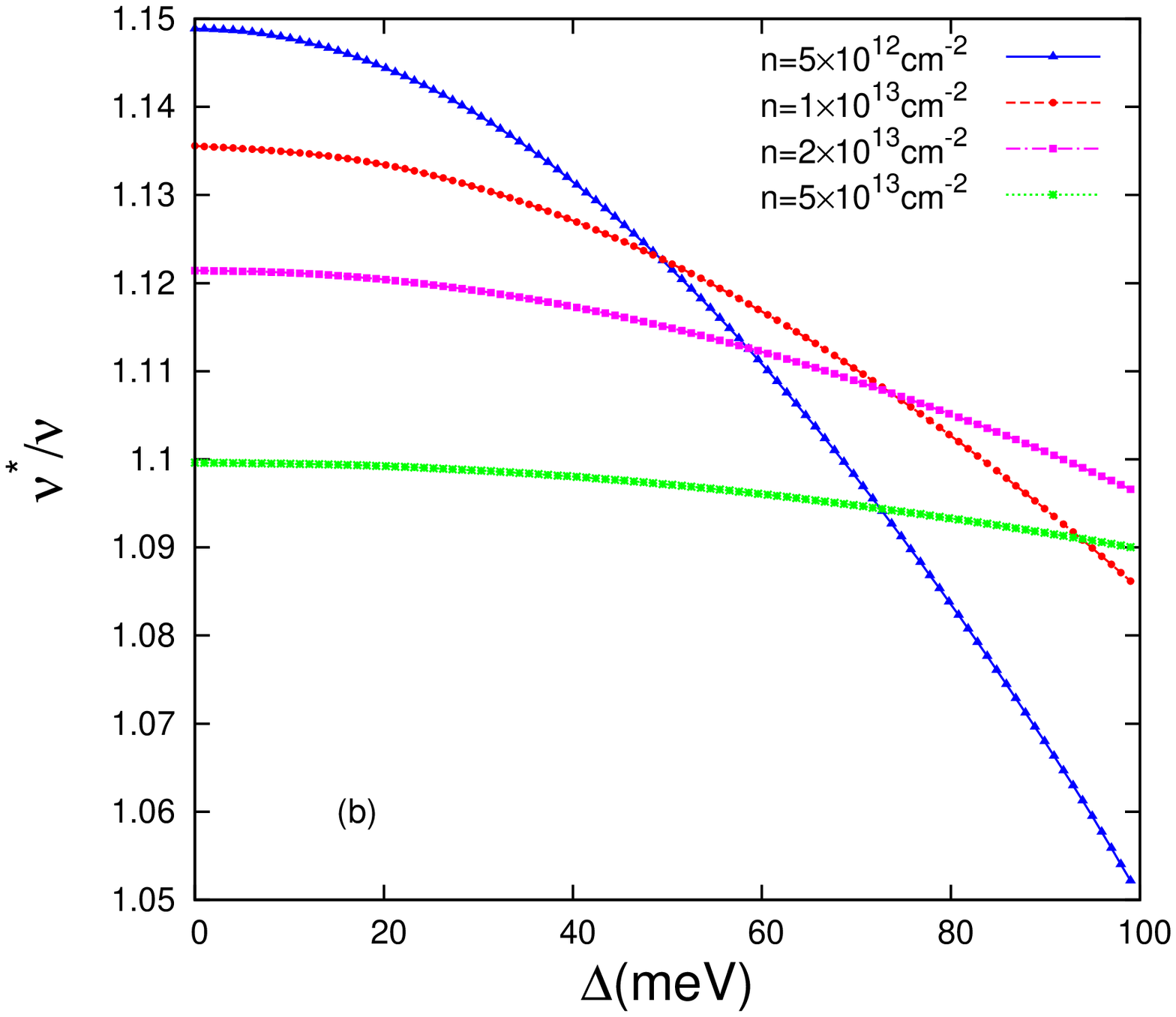}
\caption{( Color online) (a): The renormalized velocity as a function of density
for the various energy gaps at $\alpha_{gr}=1$. (b): The
renormalized velocity as a function of the energy gap for the various
densities. } \label{}
\end{figure}

\begin{figure}[t]
\includegraphics[width=8cm]{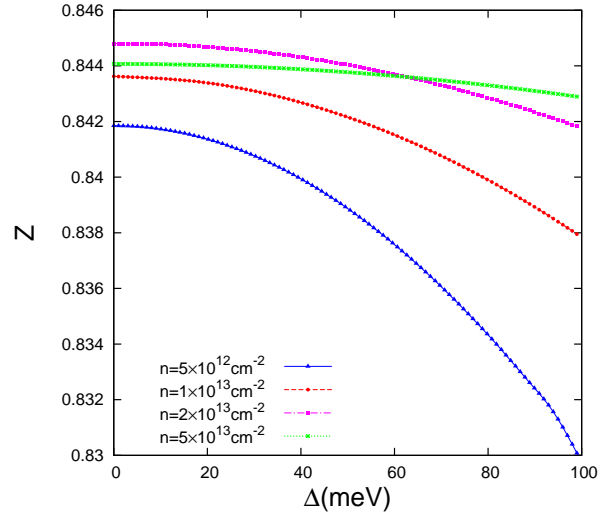}
\caption{( Color online) The renormalization constant as a function of the energy
gap for the various densities.} \label{}
\end{figure}

\begin{figure}[h]
\includegraphics[width=8cm]{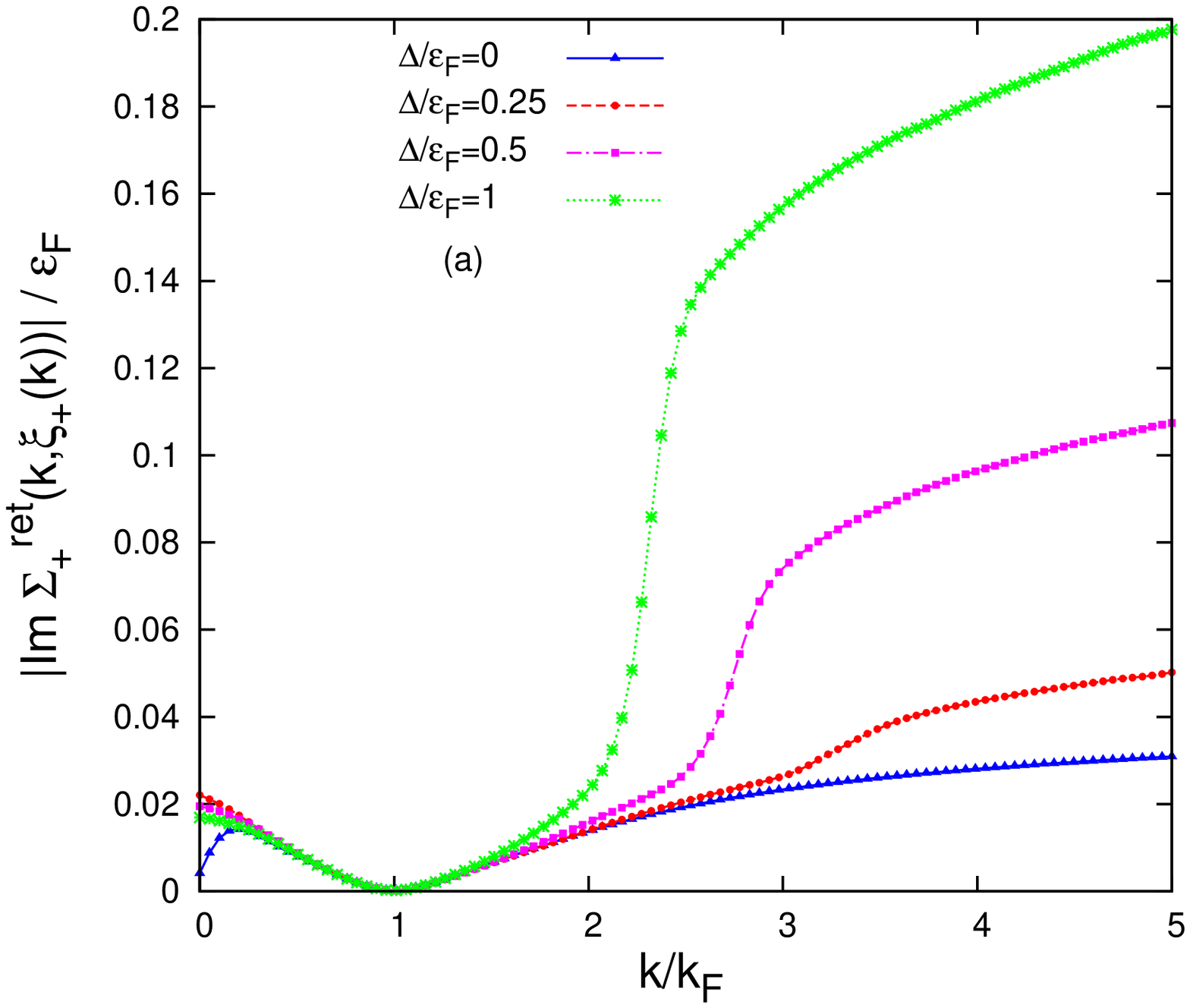}
\includegraphics[width=8cm]{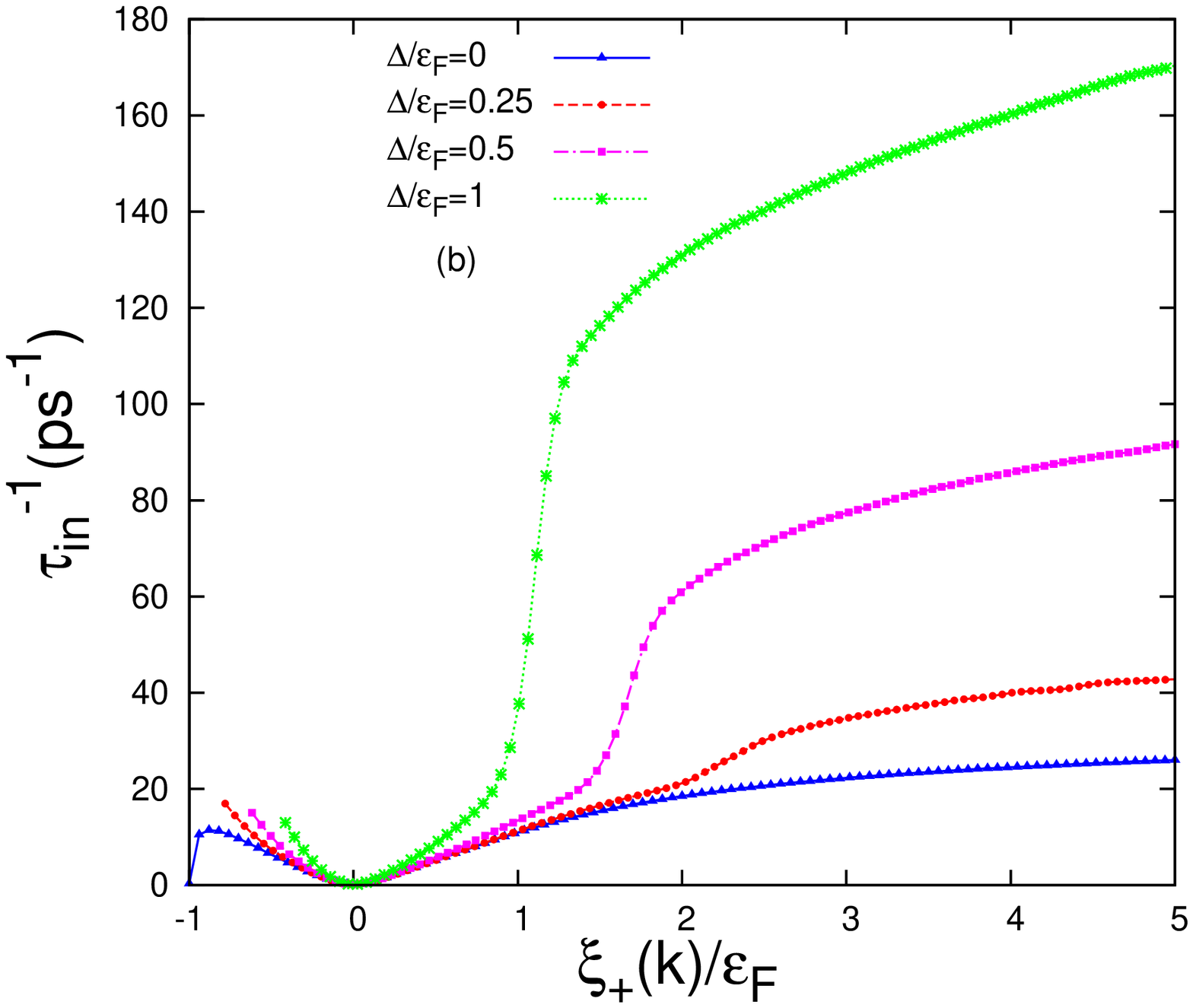}
\caption{( Color online) (a): The absolute value of the imaginary part of the
retarded self-energy on the energy shell as a function of the
wavevector for the various energy gaps; (b):
The inelastic quasiparticle lifetime($\tau_{in}$) in graphene as a function of the
on-shell energy for the various energy gaps at
$n=5\times10^{12}$cm$^{-2}$. } \label{}
\end{figure}

\begin{figure}[t]
\includegraphics[width=8cm]{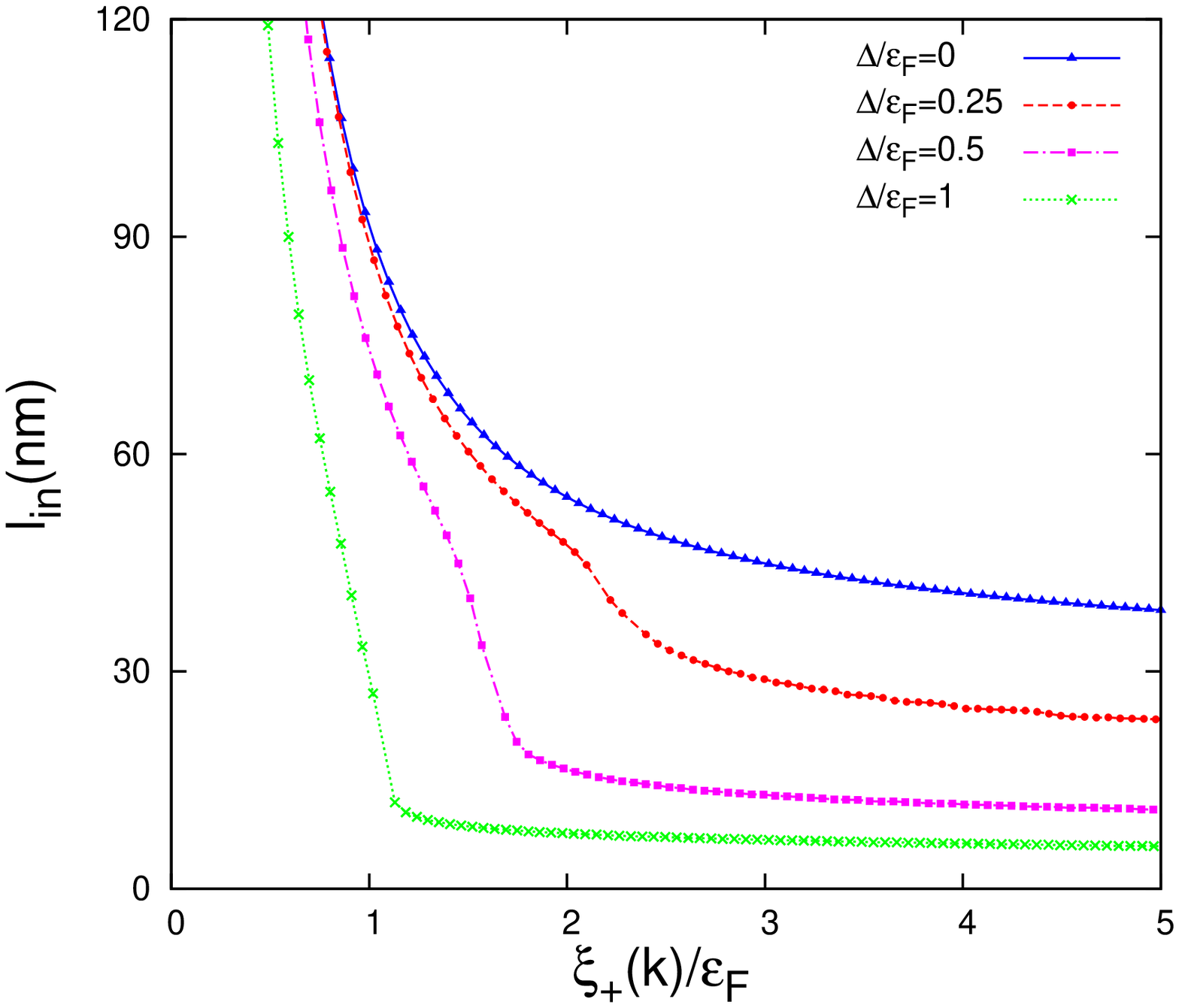}
\caption{( Color online) The
quasiparticle mean free path as a function of the on-shell energy
for the various energy gaps at $n=5\times10^{12}$cm$^{-2}$.} \label{}
\end{figure}


\begin{thebibliography}{99}
\bibitem{novoselov}
    K. S. Novoselov, A. K. Geim, S. V. Morozov, D. Jiang,
    Y. Zhang, S. V. Dubonos, I. V. Grigorieva, A. A. Firsov,
    Science {\bf 306}, 666 (2004);
    K. S. Novoselov, A. K. Geim, S. V. Morozov, D. Jiang, M. I. Katsnelson, I.
    V. Grigorieva, S. V. Dubonos and A. A. Firsov, Nature {\bf 438}, 197 (2005);
    K. S. Novoselov, D. Jiang, F. Schedin, T. J. Booth, V. V. Khotkevich, S.
    V. Morozov, and A. K. Geim, Proc. Nat. Acad. Sci. {\bf 102}, 10451 (2005)~.
\bibitem{Barth}
    Andreas Barth, Werner Marx, arXiv:0808.3320 (2008)~.
\bibitem{thermal}
    A. A. Balandin, S. Ghosh, W. Bao, I. Calizo, D. Teweldebrhan, F. Miao, C. N.
    Lau, Nano Lett., {\bf 8}, 902 (2008)~.
\bibitem{morozov}
    S.V. Morozov, K.S. Novoselov, M.I. Katsnelson, F. Schedin, D.C. Elias, J.A. Jaszczak, A.K.
    Geim, Phys. Rev. Lett. {\bf 100}, 016602 (2008);
    K. I. Bolotin, K. J. Sikes, J. Hone, H. L. Stormer, and P.
    Kim, Phys, Rev, Lett. {\bf 101}, 096802 (2008);
    Xu Du, Ivan Skachko, Anthoy Barker and Eva Y. Andrei, Nature
    Nanotech. {\bf 3}, 491 (2008);
    K.I. Bolotin, K.J. Sikes, Z. Jiang, M. Klima, G. Fudenberg, J. Hone, P. Kim, H.L.
    Stormer, Solid State Commun. {\bf 146}, 351 (2008)~.
\bibitem{eng}
    K. Eng, R. N. McFarland, and B. E. Kane, Appl. Phys. Lett. {\bf
    87}, 052106 (2005); E. H. Hwang and S. Das Sarma, Phys. Rev. B
    {\bf 75}, 073301 (2007)~.
\bibitem{dop}
    T. O. Wehling, K. S. Novoselov, S. V. Morozov, E. E. Vdovin, M. I. Katsnelson, A. K. Geim, A. I.
    Lichtenstein, Nano Lett.~{\bf 8}, 173-177 (2008); S.Y. Zhou, D.A. Siegel, A.V. Fedorov, A.
    Lanzara, Phys. Rev. Lett. {\bf 101}, 086402 (2008);
    Isabella Gierz, Christian Riedl, Ulrich Starke, Christian R. Ast, Klaus
    Kern, arXiv:0808.0621~.
\bibitem{3}
    S. Y. Zhou, G. H. Gweon, A. V. Federov, P. N. First, W. A. de
    Heer, D. H. Lee, F. Guinea, A. H. Castro Neto, and A. Lanzara,
    Nat. Mater. {\bf 6}, 770 (2007)~.
\bibitem{4}
    G. Li, A. Luican, and E. Y. Andrei, arXiv:0803.4016 (2008)~.
\bibitem{5}
    Gianluca Giovannetti, Pet A. Khomyakov, Geert Brocks, Paul J. Kelly, and Jeoroen Van den Brink,
    Phys. Rev. B {\bf 76}, 073103 (2007)~.
\bibitem{6}
    R. M. Ribeiro, N. M. R. Peres, J. Coutinho, and P. R. Briddon, Phys. Rev. B {\bf 78}, 075442 (2008)~.
\bibitem{cro}
    I. Zanella, S. Guerini, S. B. Fagan, J. Mendes Filho, and A. G. Souza
    Filho, Phys. Rev. B {\bf 77}, 073404 (2008)~.
\bibitem{spin}
    Yugui Yao, F. Ye, Xiao-Liaug Qi, Shou-Cheng Zhang and Zhang Fang, Phys. Rev. B {\bf 75}, 041401(R) (2007);
    C. L. Kane, and E. J. Mele, Phys. Rev. Lett. {\bf 95}, 226801 (2005); Hongki Min, J. E. Hill,
    N. A. Sinitsyn, B. R. Sahu, Leonard Kleinman, and A. H. MacDonald, Phys. Rev. B {\bf 74}, 165310 (2006)~.
\bibitem{size}
    Y. W. Son, M. L. Cohen, and S. G. Louie, Phys. Rev. Lett. {\bf 97}, 216803 (2006); M. Y. Han, B. Ozyilmaz,
    Y. Zhang, and P. Kim, \textit{ibid} {\bf 98}, 206805 (2007)~.
\bibitem{landau}
    L.D. Landau, Sov. Phys. JEPT, {\bf 3}, 920 (1957)~.
\bibitem{em}
    I.K. Marmorkos and S. Das Sarma, Phys. Rev. B {\bf 44}, R3451 (1991);
    J.D. Lee and B.I. Min, Phys. Rev. B {\bf 53}, 10\,988 (1996);
    H.-J. Schulze, P. Schuck, and N. Van Giai, Phys. Rev. B {\bf
    61}, 8026 (2000)~.
\bibitem{yarlagadda_1994_2}
    S. Yarlagadda and G.F. Giuliani,
    Phys. Rev. B {\bf 49}, 7887 (1994); {\bf 61}, 12556 (2000);
    C.S. Ting, T.K. Lee, and J.J. Quinn,
    Phys. Rev. Lett. {\bf 34}, 870 (1975).
\bibitem{em_bohm}
    H. M. B{\" o}hm and K. Sch{\" o}rkhuber,
    J. Phys.: Condens. Matter {\bf 12}, 2007 (2000)~.
\bibitem{em_dassarma}
    Y. Zhang and S. Das Sarma,
    Phys. Rev. B {\bf 71}, 045322 (2005);
    S. Das Sarma, V. M. Galitski, and Y. Zhang,
    Phys. Rev. B {\bf 69}, 125334 (2004)~.
\bibitem{zhang}
    Y. Zhang, V.M. Yakovenko, and S. Das Sarma,
    Phys. Rev. B {\bf 71}, 115105 (2005)~.
\bibitem{asgari}
    R. Asgari, B. Davoudi, M. Polini, Gabriele F. Giuliani, M. P.
    Tosi, and G. Vignale, Phys. rev. B {\bf 71}, 045323 (2005);  R.
    Jalabert and S. Das Sarma, Phys. Rev. B {\bf 40}, 9723 (1989)~.
\bibitem{asgari12}
    R. Asgari, B. Davoudi and B. Tanatar,
    Solid State Commun. {\bf 130}, 13 (2004).
\bibitem{tau}
    Jahan M. Dawlaty, Shriram Shivaraman, Mvs Chandrasekhar,
    Farhan Rana, Micheal G. Spencer, App. Phys. Lett {\bf 92}, 042116 (2008)~.
\bibitem{williams}
    J. R. Williams, L. DiCarlo and C. M. Marcus Since {\bf 317}, 638 (2007)~.
\bibitem{gu}
    G. Gu, S. Nie, R. M. Feenstra, R. P. Devaty, W. J. Choyke, W. K. Chan and M. G. Kane, Appl. Phys. Lett. {\bf 90}, 253507 (2007)~.

\bibitem{im}
    Marco Polini, Reza Asgari, Giovanni Borghi, Yafis Barlas, T. Pereg-Barnea, and A.H.
    MacDonald, Phys. Rev. B {\bf 77}, 081411(R) (2008); E. H.
    Hwang and S. Das Sarma, Phys. Rev. B {\bf 77}, 081412(R) (2008)~.

\bibitem{martin}
    Z. Q. Li, E. A. Henriksen, Z. Jiang, Z. Hao, M. C. Martin, P.
    Kim, H. L. Stromer, D. N. Basov, Nature Phys. {\bf 4}, 532
    (2008)~.

\bibitem{velocity}
    M. Polini, R. Asgari, Y. Barlas, T. Pereg-Barnea, A. H.
    MacDonald, Solid State Commun. {\bf 143}, 58 (2007)~.
\bibitem{Giuliani}
    G. F. Giuliani and G. Vignale, \textit{Quantum Theory of The Electron
    Liquid} (Cambridge University Press, Cambridge, England, 2005).
\bibitem{alireza}
    A. Qaiumzadeh and R. Asgari, arXiv:0807.3183 (2008)~.
\bibitem{pyatkovskiy}
    P. K. Pyatkovskiy, arXiv:0808.0931 (2008)~.

\bibitem{inelastic}
    E. H. Hwang, BenYu-Kaung Hu, and S. Das Sarma, Phys. Rev. B {\bf 76}, 115434 (2007)~.
\bibitem{asgari2}
    R. Asgari and B. Tanatar, Phys. Rev. B 65 (2002) 085311 ~.
    \bibitem{yafis}
    Y. Barlas, T. Pereg-Barnea, M. Polini, R. Asgari and A. H.
    MacDonald, Phys. Rev. Lett. 98 (2007) 236601~.
\bibitem{impurity}
    A. Qaiumzadeh, N. Arabchi, R. Asgari, Solid State Commun. {\bf
    147}, 172 (2008)~.
 \bibitem{giuliani_quinn}
    G.F. Giuliani and J. J. Quinn, Phys. Rev. B {\bf 26}, 4421 (1982)~.
 \bibitem{private}
    P. K. Pyatkovskiy, Private communication~.
\end{thebibliography}
\end{document}